\documentclass[preprint,12pt]{aastex}

\begin{document}

\title{The Effect of the Approach to Gas Disk Gravitational Instability 
on the Rapid Formation of Gas Giant Planets. II. Quadrupled Spatial 
Resolution}

\author{Alan P. Boss}
\affil{Earth \& Planets Laboratory, Carnegie Institution
for Science, 5241 Broad Branch Road, NW, Washington, DC 20015-1305}
\authoremail{aboss@carnegiescience.edu}

\begin{abstract}

 Observations support the hypothesis that gas 
disk gravitational instability might explain the formation of 
massive or wide-orbit gas giant exoplanets. The situation with regard
to Jupiter-mass exoplanets orbiting within $\sim$ 20 au is more
uncertain. Theoretical models yield divergent assessments
often attributed to the numerical handling of the gas thermodynamics.
Boss (2019) used the $\beta$ cooling approximation 
to calculate three dimensional hydrodynamical models of the evolution of
disks with initial masses of 0.091 $M_\odot$ extending from 4 to 20 au
around 1 $M_\odot$ protostars. The models considered a wide range 
(1 to 100) of $\beta$ cooling parameters and started from an initial 
minimum Toomre stability parameter of $Q_i = 2.7$ (gravitationally stable).
The disks cooled down from initial outer disk temperatures of 180 K to 
as low as 40 K as a result of the $\beta$ cooling, leading to  
fragmentation into dense clumps, which were then replaced by virtual 
protoplanets (VPs) and evolved for up to $\sim$ 500 yr. The present
models test the viability of replacing dense clumps with VPs
by quadrupling the spatial resolution of the grid once dense clumps 
form, sidestepping in most cases VP insertion. After at least $\sim$ 200 
yr of evolution, the new results compare favorably with those of 
Boss (2019): similar numbers of VPs and dense clumps form by the same 
time for the two approaches. The results imply that VP insertion can 
greatly speed disk instability calculations without sacrificing accuracy.

\end{abstract}


\section{Introduction}

 Core accretion (e.g., Mizuno 1980) has long been considered the primary
mechanism for gas giant planet formation, with the alternative of
gas disk gravitational instability (e.g., Boss 1997a) being considered
a long shot (e.g., see the review by Helled et al. 2014). 
In recent years, however, observational evidence has
begun to suggest at least a limited role for disk instability in the
formation of giant planets. A direct imaging survey found an
occurrence rate of $\sim 9 \%$ for exoplanets with masses in the range
from 5 to 13 $M_{Jup}$ at distances of 10 to 100 au around high-mass
($> 1.5 M_\odot$) stars (Nielsen et al. 2019). A compilation of direct 
imaging surveys for 344 stars found an occurrence rate of $\sim 11 \%$ 
for companions with masses in the range from 1 to 20 $M_{Jup}$ at 
distances of 5 to 5000 au (Baron et al. 2019). Both studies concluded
that the wider separation gas giants likely formed by gravitational 
instability, as is usually inferred for brown dwarf companions. In situ
formation at distances of about 160 au and 320 au, respectively, seems to
be required for the two gas giant planets orbiting the young star TYC 
8998-760-1 (Bohn et al. 2020), as only circular orbits are stable for
these exoplanets.

 Ongoing observational work has shown that the strong dependence of
exoplanet frequency on host star metallicity found by Fischer \& Valenti 
(2005) has been supplanted by a nearly symmetrical frequency distribution 
centered on solar metallicity (Figure 1), weakening the overall case
for exoplanet formation primarily by core accretion.
Santos et al. (2017), Schlaufman (2018), Narang et al. (2018),
Goda \& Matsuo (2019), and Maldonado et al. (2019) all
found that the distribution of exoplanet masses as a function of stellar
metallicity depends on the exoplanet mass. Exoplanets with masses below 
$\sim 4 M_{Jup}$ preferentially orbit high metallicity stars, 
while those with higher masses do not. Considering that core accretion
is expected to be less efficient around low metallicity stars, as a
consequence of a reduced feedstock for core assembly from planetesimals
or pebbles, these studies imply that disk instability may be primarily
responsible for the formation of the more massive exoplanets.
Furthermore, Teske et al. (2019) did not find the clear correlation
between host star metallicity and giant exoplanet excess metallicity
that might be expected for formation by core accretion.

 Core accretion has long had difficulty with forming gas giants around
low mass, M dwarf stars (e.g., Laughlin et al. 2004), as opposed to the
success of disk instability (e.g., Boss 2006). More recent work has 
confirmed the core accretion problem for low mass stars (Miguel et al. 2020),
while M dwarfs still appear capable of forming gas giants by disk 
instability, provided their
disks are massive enough (Mercer \& Stamatellos 2020). The discovery
of a gas giant planet orbiting the M5.5 dwarf star GJ 3512 has
therefore been attributed to it having been formed by disk instability
rather than by core accretion (Morales et al. 2019). High mass
stars also appear to favor fragmentation as a formation process
for their more massive companions (Cadman et al. 2020).

 FU Ori-type stars undergo rapid luminosity increases that are 
consistently understood to be the result of a phase of gravitational
instability and fragmentation (e.g., Takami et al. 2018). ALMA was 
used by Cieza et al. (2018) to show that FU Ori-type disks have radii 
less than 20 to 40 au and masses of 0.08 to 0.6 $M_\odot$, which
are expected to be at least marginally gravitationally unstable.
ALMA observations of the HL Tau disk have been used to interpret a 
disk gap as being caused by a gas giant planet in the process of 
formation by disk instability (Booth \& Ilee 2020). ALMA disk gaps
have been interpreted to be caused by embedded protoplanets with
masses in the range of $\sim$ 0.1 to $\sim$ 10 $M_{Jup}$ at distances 
from $\sim$ 10 to $\sim$ 100 au (Zhang et al. 2018). ALMA imaging of
the young stellar object BHB1 (Alves et al. 2020), with a disk gap, 
suggests that a massive gas giant planet has formed at the same time 
that the protostar is still accreting, i.e., at a very early age, less than
1 Myr. Four annular structures have been found by ALMA in a protostellar
disk less than 0.5 Myr old (Segura-Cox et al. 2020), again implying 
rapid formation of protoplanets at large distances, presumably 
implicating disk gravitational instability. ALMA has also
been used to study the spiral arms surrounding the AFGL 4176 O-type star,
finding that the outer disk is cool enough to be gravitationally
unstable and is likely to fragment into a companion object (Johnston 
et al. 2020). In fact, outer disk temperatures for the HD 163296 
protoplanetary disk drop to $\sim$ 18 K in the midplane beyond 
100 au (Dullemond et al. 2020). While disks with spiral arms appear
to be rare, this may be more a result of spiral arm suppression by
migrating giant planets than of insufficient mass for the disks to be
gravitationally unstable (Rowther et al. 2020). When spiral arms do
appear, they need not be caused by embedded planets, but rather
may be direct manifestations of disk gas gravitational instability
(Chen et al. 2021; Xie et al. 2021).

 Boss (2017) adopted the $\beta$ cooling approximation for disk 
thermodynamics to study instabilities in disks with a wide range of 
$\beta$ values (1 to 100), starting with initial disks ranging from 
gravitationally unstable (Toomre $Q_{min} = 1.3$) to gravitationally 
stable ($Q_{min} = 2.7$), finding that fragmentation could occur 
for all $\beta$ values so long as the disks were initially close
enough to instability ($Q_{min} < 2$). Boss (2019) then addressed
the question of evolving toward such an unstable configuration from
an initially gravitationally stable configuration that is being cooled. 
Starting with $Q_{min} = 2.7$ disks, but with the minimum disk temperature 
allowed to cool down to 40 K (equivalent to the outer disk temperature of 
models with initial $Q_{min} = 1.3$), Boss (2019) found that fragmentation
could still occur for the same wide range of $\beta$ values.   

 The present paper presents a suite of disk instability models that 
are variations on the $\beta$ cooling models of Boss (2019),
now computed with as much as quadrupled spatial resolution of the 
spherical coordinate grid to avoid the need for the
insertion of virtual protoplanets (VPs). The Boss (2019) models
were restricted to a maximum of 200 radial ($r$) 
and 1024 azimuthal ($\phi$) grid
points prior to VP insertion, whereas the new models double both
of these maxima, up to 400 in radius and 2048 in azimuth (the 
vertical resolution was unchanged). These are the
highest spatial resolution grid models ever run with the EDTONS code.
However, this enhanced spatial resolution comes only with the severe 
penalty of running potentially eight times slower than the highest 
resolution models presented in Boss (2019). The factor of eight
results from having the Courant time step halved and having four 
times as many grid points to compute at each time step. The new models
are thus designed to learn to what extent the previous models might be
considered to be close to having converged to the ideal of the
continuum limit of infinite spatial resolution. 

\section{Numerical Methods and Initial Conditions}

 The code is the same as that used in many previous studies of disk 
instability by the author (e.g., Boss 2005, 2006, 2007, 2013, 2017, 2019).
The EDTONS code solves the three-dimensional 
equations of hydrodynamics and the Poisson equation for the gravitational 
potential, with second-order-accuracy in both space and time, on a spherical
coordinate grid (see Boss \& Myhill 1992). The van Leer type hydrodynamic 
fluxes have been modified to improve stability (Boss 1997b). Numerous 
tests of the code are presented in Boss \& Myhill (1992).

 The energy equation is solved explicitly in conservation law form, 
as are the five other hydrodynamic equations. As in Boss (2017),
the numerical code solves the specific internal energy $E$ equation 
(Boss \& Myhill 1992): 

$${\partial (\rho E) \over \partial t} + \nabla \cdot (\rho E {\bf v}) =
- p \nabla \cdot {\bf v} + L, $$

\noindent
where $\rho$ is the gas density, $t$ is time, ${\bf v}$ is the velocity, 
$p$ is the gas pressure, and $L$ is the time rate of change of energy 
per unit volume, which is normally taken to be that due to the transfer 
of energy by radiation in the diffusion approximation. $L$ is defined
in terms of the $\beta$ cooling formula (Gammie 2001), as follows. With
$\beta = t_{cool} \Omega$, $t_{cool}$ is defined as the ratio of the specific
internal energy to the time rate of change of the specific internal
energy. $L$ is thus defined to be:

$$L = - {\rho E \Omega \over \beta}, $$

\noindent
where $\Omega$ is the local angular velocity of the gas. 
$L$ is always negative with this formulation, i.e., only cooling
is permitted.

Detailed equations of state for the gas (primarily molecular hydrogen) and 
dust grain opacities are employed in the models; an updated 
energy equation of state is described by Boss (2007). The central 
protostar is effectively forced to wobble in order to preserve the location 
of the center of mass of the entire system (Boss 1998), which is 
accomplished by altering the apparent location of the point mass source of 
the star's gravitational potential in order to balance the center of 
mass of the disk. No artificial viscosity is used in the models.

 The numerical grid has $N_r = 100$, 200, or 
400 uniformly spaced radial grid points, $N_\theta = 23$ theta grid points,
distributed from $\pi/2 \ge \theta \ge 0$ and compressed toward
the disk midplane, and $N_\phi = 512$, 1024, or 2048 uniformly spaced
azimuthal grid points. The radial grid extends from 4 to 20 au,
with disk gas flowing inside 4 au being added to the central 
protostar, whereas that reaching the outermost shell at 20 AU loses
its outward radial momentum but remains on the active hydrodynamical grid.
The $\theta$ grid points are compressed into the midplane to ensure 
The gravitational potential is obtained through a
spherical harmonic expansion, including terms up to $N_{Ylm} = 48$ 
for all spatial resolutions.

 The $r$ and $\phi$ numerical resolution is doubled and then
quadrupled when needed to avoid violating the Jeans length 
(e.g., Boss et al. 2000) and Toomre length criteria (Nelson 2006). 
As in Boss (2017, 2019), if either criterion is violated,
the calculation stops, and the data from a time step prior to the 
criterion violation is used to double the spatial resolution in the 
relevant direction by dividing each cell into half while conserving 
mass and momentum. Here, however, the models can be doubled 
again to as high a
spatial resolution as $N_r = 400$ and $N_\phi = 2048$ if needed.
Even with this quadrupled spatial resolution, in one model (beq3)  
dense clumps formed that violated the Jeans or Toomre length
criteria at their density maxima. In that case, the maximum density
cell is again drained of 90\% of its mass and momentum, which is then inserted 
into a virtual protoplanet (VP, Boss 2005), as in Boss (2017, 2019).
The VPs orbit in the disk midplane, 
subject to the gravitational forces of the disk gas, the central protostar, 
and any other VPs, while the disk gas is subject to the gravity of the VPs. 
VPs gain mass at the rate (Boss 2005, 2013) given by the 
Bondi-Hoyle-Lyttleton (BHL) formula (e.g., Ruffert \& Arnett 1994), as
well as the angular momentum of any accreted disk gas. 
As in Boss (2017, 2019), VPs that reach the the inner or outer boundaries 
are simply tallied and removed from the calculation. 

  In the Boss (2017, 2019) models, the initial gas disk density distribution 
is that of an adiabatic, self-gravitating, thick disk with a mass
of $M_d = 0.091 M_\odot$, in near-Keplerian 
rotation around a solar mass protostar with $M_s = 1.0 M_\odot$ (Boss 1993).
The initial outer disk temperature was set to 180 K for all models,
yielding an initial minimum value of the Toomre (1964) $Q$ gravitational 
stability parameter of 2.7, i.e., gravitationally stable, though
the disks were allowed to cool down to as low as 40 K, as in Boss (2019).
As in Boss (2017, 2019), the values of $\beta$ that were explored
(see Table 1) were 1, 3, 10, 20, 30, 40, 50, and 100.

 Gammie (2001) proposed that the outcome of a gas disk gravitational
instability would depend on the parameter $\beta$, and suggested a 
critical value for fragmentation of $\beta_{cr} = 3$. 
As a result, a critical value of $\beta_{cr} = 3$ became the standard 
for predicting the outcome of the fragmentation process in protoplanetary disks.
However, subsequent work by many groups (as summarized in Boss 2017, 2019) 
has questioned whether or not the value of $\beta_{cr} = 3$ is a correct indicator
for disk fragmentation, with other estimates of the true value ranging from 
$\beta_{cr} = 10$ to $\beta_{cr} = 30$, depending on the details of the 
numerical code used, such as numerical resolution and the artificial viscosity 
for smoothed-particle hydrodynamics (SPH) codes. 
More recently, Deng et al. (2017) found evidence for 
$\beta_{cr} = 3$ for a novel meshless finite mass (MFM) code, but not 
for an SPH code, the latter apparently a result of artificial viscosity. 
Baehr et al. (2017) used local three-dimensional disk simulations to 
find $\beta_{cr} \sim 3$. Mercer et al. (2018) showed which of two 
approximate radiative transfer procedures is more accurate for 
protostellar disks, and demonstrated that the effective value of $\beta$
in such disks could vary from $\sim$ 0.1 to $\sim$ 200, i.e., a single, constant value 
of $\beta$ may not be capable of representing the full range of physical 
conditions in gravitationally unstable disks.

 Boss (2017) discussed the problems of radiative
transfer and cooling in disk instability calculations and the utility of
the $\beta$ cooling approximation in sidestepping some of these issues. 
Boss (2001) calculated the first disk instability models with 3D radiative transfer 
hydrodynamics (RHD), finding that disk fragmentation was possible.
Durisen et al. (2007) summarized much of the detailed work on 3D radiative
transfer models, both from the side of models where clump formation 
occurred and from models where clump formation was less robust.
While many numerical factors come to play in these calculations,
e.g., grid resolutions for finite difference (FD) codes and smoothing lengths
for  SPH codes (e.g., Mayer et al. 2007), 
flux limiters for radiative transfer in the diffusion approximation, and  
accuracy of the gravitational potential solver, the key issue was determined 
to be whether a protoplanetary disk could remain sufficiently cold for
spiral arms to form self-gravitating clumps that could
contract toward planetary densities.
Steiman-Cameron et al. (2013) studied the effect of spatial 
resolution on cooling times in RHD models, finding convergence for optically 
thick, inner regions, but not for optically thin, outer regions.
Tsukamoto et al. (2015) used 3D RHD models to 
follow the formation of disks, starting from collapsing molecular cloud 
cores, finding that radiative heating from the interstellar medium 
could have a significant effect on the fragmentation process.

 Note that as in the previous disk instability models in this series,
magnetic fields can be neglected because of the low ionization levels 
in the optically thick disk midplanes (Gammie 1996; Boss 2005), where 
marginally gravitationally unstable disk dynamics occurs, at least for massive
disks of the sizes (radii from 4 au to 20 au) considered here. Magnetic
fields are expected to be of importance close to the protostar,
at the disk surfaces, and at large distances from the protostar,
especially in regions with strong EUV/FUV radiation from nearby O stars.

\section{Results}

 The new models were started from the last saved time step of
the model with corresponding $\beta$ in Boss (2019) before VPs 
were inserted, i.e., when the grids had only been doubled in
$r$ and $\phi$, but not yet quadrupled. Table 1 lists 
the key results for all of the models: the final times reached, 
the number of VPs and of strongly (first number) or strongly and
weakly (second number, when two numbers are given)
gravitationally bound clumps present at the final time,
the sum of those two ($N_{VP}$ + $N_{clumps}$ = $N_{total}$),
and the number of VPs in Boss (2019) at the two time steps closest
to the final time of the new models (a single number is given when 
both time steps have the same number of VPs). The number of clumps
($N_{clumps}$) was assessed by searching for dense regions with
densities greater than $10^{-10}$ g cm$^{-3}$. For clumps of
this density or higher, the free fall time is 6.7 yrs or less,
considerably less than the orbital periods, implying that such clumps
might be able to survive and contract to form gaseous protoplanets,
provided there was sufficient spatial resolution. The orbital period 
of the disk gas at the inner edge (4 au) is 8.0 yr and 91 yrs at 
the outer edge (20 au). The final times reached ranged from 205 yrs 
to 326 yrs, indicating that the models spanned time periods long 
enough for many revolutions in the inner disk and multiple 
revolutions in the outer disk. The models required about 2.5 years to 
compute, with each model running on a separate, single core of the 
Carnegie memex cluster at Stanford University. 

 The previous models with $\beta$ cooling (Boss 2017, 2019) all began
from axisymmetric initial disks, and followed the growth of nonaxisymmetry
as a result of disk self-gravitational instability, i.e., the formation and evolution
of spiral arms and clumps, as shown in Figures 2 and 3 of Boss (2019).
The present models all begin from the corresponding $\beta$ models of
Boss (2019), just before the densest clumps were replaced with VPs,
and so began highly nonaxisymmetric and clumpy.  Figures 2 and 3
display the density and temperature distributions in the disk midplane
for the eight new models with higher spatial resolution, showing
the familiar proliferation of spiral arms and segments in both fields.
As expected, the highest disk temperatures occur in the regions of
highest disk density, where $\beta$ cooling is struggling to overcome
the compressional heating of the disk gas in the clumps and arms.

 In order to avoid violating the Jeans and Toomre length criteria
during the evolutions, models beq1 through beq6 all required that
the radial and azimuthal grids be doubled compared to the highest
resolution used in the Boss (2019) models, that is, increased to the
maximum values of $N_r = 400$ and $N_\phi = 2048$. In spite of these
increases, model beq3 twice required the insertion of a VP in order to 
avoid violating the two length constraints, though only one VP was still
active at the end of the calculation (Table 1). Models beq7 and
beq8, on the other hand, only needed to be refined in radius in order
to avoid the length criteria, and so reached a highest spatial resolution
of  $N_r = 400$ and $N_\phi = 1024.$ Given the slower  $\beta$ cooling
in models beq7 and beq8 than in the six lower $\beta$ models, this
is to be expected. As in all the models in this series,
the vertical resolution was fixed, with the theta grid cells compressed
around the disk midplane for maximum vertical spatial resolution
of the most dynamically evolving regions of the disk.

As noted above, Table 1 lists a range for the number of clumps
present in the new models at the final time, with the range being the
number of strongly or strongly and weakly gravitationally bound 
clumps, as defined below and in Table 2. Table 2 displays the clump 
properties for the eight new models upon which the numbers of 
clumps in Table 1 were assessed. The clumps are classified as either 
unbound (U), weakly (W), or strongly (S) self-gravitating, depending
on whether the clump mass is less than (U), greater than (W), or more 
than 1.5 times (S) the Jeans mass for self-gravitational collapse.
The Jeans mass is the mass of a sphere of uniform density gas
with a radius equal to the Jeans length, where the Jeans length is
the critical wavelength for self-gravitational collapse of an isothermal
gas (e.g., Boss 1997b, 2005). In cgs units, the Jeans mass is given
by $1.3 \times 10^{23} (T/\mu)^{3/2}\rho^{-1/2}$, where $T$ is the gas
temperature, $\rho$ the gas density, and $\mu$ the gas mean molecular 
weight. Clumps with masses exceeding the Jeans mass are capable of
self-gravitational collapse on a time scale given by the free fall time.
As previously noted, all of these clumps have a maximum density
higher than $10^{-10}$ g cm$^{-3}$ (Table 2), implying collapse times
shorter than a free fall time of 6.7 yr and thus shorter than their orbital
periods.

 An inspection of the last two columns in Table 1 shows that the total
number of clumps and VPs formed in the present models ($N_{total}$)
tracks reasonably well with the number of VPs ($N_{VP-2019}$) in
the corresponding Boss (2019) models. The numbers in both
cases drop from about 4 to 6 down to 1 to 2 as the $\beta$
parameter increases from 1 to 100, as to be expected as the cooling
rate decreases. This result implies that the quadrupled spatial
resolution of the present models, with the consequent major increase
in computational times, does not appear to produce results that
differ greatly from those obtained in the same disk model when the
virtual protoplanet technique is employed to represent the highest
density clumps. 

The clump masses range from 0.30 $M_{Jup}$ to 2.3
$M_{Jup}$, i.e., from roughly a Saturn-mass to twice a Jupiter-mass.
Table 3 lists the orbital parameters for the clumps displayed in Tables
1 and 2, based on their masses and midplane velocities at the final
times shown in Table 1. Note that the maximum clump densities always
occur in the midplane, except for a few clumps where the maximum occurs
in the first cell above above the midplane, so the orbits summarized in
Table 3 are all assumed to be restricted to the midplane, i.e., zero 
inclination.

 Figure 4 plots the clump masses for the present models as a function
of semi-major axis, as well as the VPs from Boss (2019) at the data dump
closest in time to the final times shown in Table 1. Note that the one VP 
that formed and survived in the present models (Table 1: model beq3) had a mass of 
only $4.3 \times 10^{-3}$ $M_{Jup}$ and is not plotted in Figures 4 and 5. 
Also plotted in Figure 4 are the known exoplanets with masses and semi-major
axes in the ranges of 0.1 $M_{Jup}$ to 5 $M_{Jup}$  and 4 au to 20 au, 
respectively. Figure 4 shows that the clumps formed primarily with semi-major
axes in the range of 8 au to 16 au, whereas the VPs from Boss (2019) at 
those times orbited with semi-major axes between 6 au and 15 au. This slight
discrepancy is due to the fact that once the VPs form, they orbit 
chaotically around the disk but with a tendency to move inward to
smaller semi-major axes. This can be seen in Figure 5, where the VPs
from Boss (2019) are shown not just at the same time as the clumps
in Table 1, but at about 40 different data dumps, most of which show
evolutionary times later than those plotted in Figure 4. Figure 5 shows that
the VPs evolve to somewhat smaller semi-major axes on a time scale of a few hundred
years (the Boss 2019 models ran as long as 462 yrs), and to some extent
that inward evolution is responsible for the slight inward shift in semi-major
axes compared to the clumps in Figure 4. The VPs also gain mass by
BHL accretion rapidly over this time period, accounting for the much
larger VP masses seen in Figure 5 compared to Figure 4, which is closer
in time to when each VP was formed. The differing clump and VPs 
masses shown in Figure 4 are a result of the newly formed VPs having
their mass derived solely from the single grid cell where the maximum
density occurs, whereas the clump masses are derived by summing
all the cells adjacent to the maximum density cell with densities no less
than 0.1 that of the maximum density cell.
 
 Figure 6 shows the semi-major axes and eccentricities for the VPs
at multiple times and the clumps at their final time, again showing the
inward migration of the VPs, as well as their eccentricities, which
compare favorably in distribution to those of the clumps: most
eccentricities are in the range of 0 to 0.35, but in both cases a few
are as high as 0.9. Again, the quadrupled resolution clumps appear to 
support the VP model results in a broad sense.
 
 While the goal of the present paper is to compare the quadrupled spatial
resolution models with the VP models of Boss (2019), Figures 4 and 5 show
that the observed exoplanets tend to have significantly higher
masses and smaller semi-major axes than the clumps, though the agreement
with the VPs is considerably better (see Table 2 in Boss 2019). Given
that the observational biases for the known exoplanets favor large masses
and small semi-major axes, the implication of the Boss (2019) and
present models is that there may be a significant population of Jupiter-mass
exoplanets orbiting from about 5 au to 15 au, awaiting discovery by future
microlensing and direct imaging surveys (i.e., the Roman Space Telescope, RST).
Figure 7 depicts the $\sim 100$ confirmed microlensing exoplanet discoveries to date,
showing that a number of Jovian-mass exoplanets with semi-major axes
between 5 au and 20 au have been detected by ground-based microlensing
surveys, implying that they will be found in even greater numbers
by the RST microlensing survey.
  
\subsection{Discussion}

Jin et al. (2020) undertook a semi-analytical model of the collapse of dense
cloud cores to form protostars and protostellar disks. They found that disk
fragmentation was most likely to occur at distances of 20 au to 200 au, 
though also at 200 au to 450 au to a reduced extent, with fragment masses
expected primarily in the range of 3-35 $M_{Jup}$. Their disk model consists
of a diffusion equation solution for the evolution of the gas surface density
in an infinitely thin, one-dimensional (axisymmetric) disk, with fragmentation
assumed to occur where the Toomre $Q$ value drops below 1.4. In their models,
the Toomre $Q$ does not fall below 1.4 inside of 20 au, and so they do not
predict any fragmentation to occur inside 20 au. Their models extend in time
for about 1 Myr, by which time the typical disk extends well beyond 1000 au.
In contrast, the present fully three dimensional gas hydrodynamics models 
do not attempt to follow the initial cloud core collapse phase (e.g., Boss 1982),
but rather begin after protostar collapse has largely completed and left behind
a 20 au radius, protoplanetary disk with an initial minimum Toomre $Q$ value 
of 2.7, stable with respect to fragmentation, and then follow the disk evolutions 
as they are allowed to cool.

\section{Conclusions}

 These models have shown that increasingly higher spatial resolution models
of the evolution of gravitationally unstable protoplanetary disks continue to 
support the possibility of the formation of self-gravitating clumps inside 
distances of 20 au from a solar-mass protostar. If these clumps are able to
contract and survive their subsequent orbital evolution, they might imply
the existence of a population of roughly Jupiter-mass exoplanets orbiting
at distances that have not yet been thoroughly sampled by ground- or 
space-based telescopic surveys. The Roman Space Telescope, scheduled
for launch in 2025, may be able to detect this largely unseen population
through its gravitational microlensing survey, and perhaps may be able
to perform direct imaging of a few with its coronagraphic instrument (CGI).

 The fact that the quadrupled spatial resolution models produced results
in terms of the number of clumps and of their initial orbital properties
comparable to those of the virtual protoplanet (VP) models of Boss (2019) implies
that the VP technique should be considered as a viable means for exploring
a larger region of protoplanetary disk parameter space, as it yields 
comparable results with considerably lower computational burden than
the quadrupled spatial resolution models presented here.

 Clearly the utility of the $\beta$ cooling approximation needs to be
further explored by comparison of the results of 3D RHD calculations
with the results of 3D $\beta$ cooling models. A large set of such models
is currently being calculated with the EDTONS code and will be the
subject of a future publication. These new RHD models will also
include quadrupled spatial resolution, in an attempt to test to what
extent the models have approached convergence to the continuum
limit, i.e., infinite spatial resolution.

 The computations were performed on the Carnegie Institution memex computer
cluster (hpc.carnegiescience.edu) with the support of the Carnegie Scientific
Computing Committee. I thank Floyd Fayton for his invaluable assistance with 
the use of memex, and the referee for suggesting several additions to the
manuscript.

\clearpage

\begin{deluxetable}{lccccccc}
\tablecaption{Results for the new models with varied $\beta$ cooling
and quadrupled spatial resolution, showing the number of VPs and of 
clumps (both weakly and strongly gravitationally bound) at the final time,
the sum of those two ($N_{VP}$ + $N_{clumps}$ = $N_{total}$),
and the number of VPs in Boss (2019) at the two data dumps closest
to the final time of the new models.}
\label{tbl-1}
\tablewidth{0pt}
\tablehead{\colhead{Model} 
& \colhead{$\beta$}
& \colhead{final time (yrs)} 
& \colhead{$N_{VP}$} 
& \colhead{$N_{clumps}$} 
& \colhead{$N_{total}$}
& \colhead{$N_{VP-2019}$}}
\startdata

beq1   &    1   & 210. & 0 & 4 & 4 & 5-6 \\                              

beq2   &    3   & 206. & 0 & 4-6 & 4-6 & 4-5 \\

beq3   &   10   & 223. & 1 & 3-5 & 4-6 & 3-4 \\
 
beq4   &   20   & 267. & 0 & 0-1 & 0-1 & 2 \\

beq5   &   30   & 205. & 0 & 1-2 & 1-2 & 3 \\

beq6   &   40   & 245. & 0 & 0-3 & 0-3 & 1 \\

beq7   &   50   & 306. & 0 & 1-2 & 1-2 & 1 \\
 
beq8   &  100  & 326. & 0 & 1-2 & 1-2 & 2 \\
   
\enddata
\end{deluxetable}

\clearpage

\begin{deluxetable}{lccccccc}
\tablecaption{Results for the new models,
showing the estimated properties for the densest clumps present at the
final times of the models, classified as either unbound (U), weakly (W),
or strongly (S) self-gravitating, based on whether the clump mass
is less than (U), greater than (W), or more than 1.5 times (S) 
the Jeans mass for self-gravitational collapse, with 
implications for clump survival. }
\label{tbl-2}
\tablewidth{0pt}
\tablehead{\colhead{Model} 
& \colhead{$\beta$}
& \colhead{clump \#} 
& \colhead{$\rho_{max}$ (g cm$^{-3}$)} 
& \colhead{$M_{clump}$/$M_{Jup}$} 
& \colhead{$T_{average}$ (K)} 
& \colhead{$M_{Jeans}$/$M_{Jup}$}
& \colhead{status}}
\startdata

beq1   &    1    & 1 &  4.2e-10  &  0.51 & 106. & 1.7    &  U  \\                              
beq1   &    1    & 2 &  5.6e-10  &  2.3   & 44.   &  0.44 &  S  \\  
beq1   &    1    & 3 &  5.7e-10  &  0.83 & 47.   &  0.52 &  S  \\  
beq1   &    1    & 4 &  9.6e-10  &  1.0   & 43.   &  0.34 &  S  \\  
beq1   &    1    & 5 &  2.5e-9    &  0.94 & 42.   &  0.22 &  S  \\  

beq2   &    3    & 1 &  2.4e-9 & 1.0 & 102. & 0.88 & W  \\                              
beq2   &    3    & 2 &  1.6e-9 & 1.6 & 92. & 0.81  & S  \\  
beq2   &    3    & 3 &  1.6e-9 & 1.4 & 51. & 0.39  & S  \\  
beq2   &    3    & 4 &  8.1e-10 & 1.2 & 42. & 0.38  & S \\  
beq2   &    3    & 5 &  4.8e-10 & 1.4 & 42. & 0.48  & S \\  
beq2   &    3    & 6 &   5.3e-10 & 0.71 & 44. & 0.48  & W \\ 

beq3   &    10   & 1 & 6.3e-9  & 0.91 & 86. & 0.41  & S  \\                              
beq3   &    10   & 2 & 9.4e-10 & 0.52  & 41. & 0.32  & S  \\  
beq3   &    10   & 3 & 4.9e-10 & 1.1 & 44. & 0.50  &  S \\  
beq3   &    10   & 4 & 7.9e-10 & 0.47 & 45. & 0.44  & W   \\  
beq3   &    10   & 5 & 1.1e-9 & 0.32 & 115. & 1.5  & U   \\  
beq3   &    10   & 6 & 3.4e-10 & 0.85 & 52. & 0.76  & W  \\ 

beq4   &    20   & 1 & 6.2e-10  & 0.48 & 56. & 0.75  & U  \\                              
beq4   &    20   & 2 & 3.7e-10  & 0.85 & 54. & 0.71  & W  \\  
beq4   &    20   & 3 & 1.3e-9 & 0.53 & 69. & 0.64  & U  \\  
beq4   &    20   & 4 &  3.7e-10 & 0.33 & 53. & 0.74  & U   \\  
beq4   &    20   & 5 &  2.9e-10 & 0.78 & 57. & 0.97  &  U  \\  

beq5   &    30   & 1 & 1.0e-9  & 0.73 & 45. & 0.39  & S  \\                              
beq5   &    30   & 2 &  9.5e-10 & 0.51 & 52. & 0.47  & W  \\  
beq5   &    30   & 3 &  1.2e-9 & 0.30 & 54. & 0.41  & U  \\  
beq5   &    30   & 4 &   2.2e-10 & 0.55 & 47. & 0.72  & U   \\  
beq5   &    30   & 5 &  2.4e-10 & 0.48 & 53. &  0.80 &  U  \\  

beq6   &    40   & 1 & 2.0e-9  & 1.1 & 99. & 0.91  & W  \\                              
beq6   &    40   & 2 &  2.4e-10 & 0.45 & 48. & 0.77  & U  \\  
beq6   &    40   & 3 &  5.9e-10 & 0.72 & 65. &  0.79 &  U \\  
beq6   &    40   & 4 &  1.0e-9 & 0.73 & 67. & 0.72  & W   \\  
beq6   &    40   & 5 &  3.7e-10 & 1.1 & 57. & 0.89  & W   \\  

beq7   &    50   & 1 &  3.1e-10 & 0.88 & 48. & 0.73  & W  \\                              
beq7   &    50   & 2 &  6.5e-10 & 0.88 & 55. & 0.59  & S  \\  
beq7   &    50   & 3 &  2.6e-10 & 0.47 & 47. & 0.76  & U  \\  

beq8   &    100   & 1 & 2.8e-10  & 1.2 & 48. & 0.71  &  S \\                              
beq8   &    100   & 2 & 3.0e-10  & 0.66 & 63. & 0.92  & U  \\  
beq8   &    100   & 3 & 2.3e-10  & 0.51 & 48. &  0.76 & U  \\  
beq8   &    100   & 4 & 6.7e-10  & 1.0 & 60. &  0.71 &  U  \\  
  
\enddata
\end{deluxetable}

\clearpage

\begin{deluxetable}{lccccc}
\tablecaption{Orbital parameters for the gravitationally bound
clumps, i.e., those with a status of W or S in Table 2, at the final
times listed in Table 1.}
\label{tbl-3}
\tablewidth{0pt}
\tablehead{\colhead{Model} 
& \colhead{$\beta$}
& \colhead{clump \#} 
& \colhead{semimajor axis (au)}
& \colhead{eccentricity}
& \colhead{status}}
\startdata
                         
beq1   &    1    & 2 & 13.85  & 0.078 & S \\  
beq1   &    1    & 3 &  9.97 & 0.061 & S \\  
beq1   &    1    & 4 &  10.92 & 0.071 & S \\  
beq1   &    1    & 5 &  12.17 & 0.011 &  S  \\  

beq2   &    3    & 1 & 8.35  & 0.072 & W  \\                              
beq2   &    3    & 2 &  8.33 & 0.076 & S  \\  
beq2   &    3    & 3 &  12.95 & 0.064 & S  \\  
beq2   &    3    & 4 & 11.93 & 0.045  & S \\  
beq2   &    3    & 5 & 12.97 & 0.054 & S \\  
beq2   &    3    & 6 &  10.06 & 0.066 & W \\ 

beq3   &    10   & 1 & 10.42 & 0.856 & S  \\                              
beq3   &    10   & 2 & 11.71 & 0.072 & S  \\  
beq3   &    10   & 3 & 18.09 & 0.256 &  S \\  
beq3   &    10   & 4 & 14.99 & 0.325 & W   \\  
beq3   &    10   & 6 & 13.81 & 0.142 & W  \\ 
                 
beq4   &    20   & 2 & 11.42 & 0.062 & W  \\  

beq5   &    30   & 1 & 14.11 & 0.066 & S  \\                              
beq5   &    30   & 2 & 9.88  & 0.023 & W  \\  

beq6   &    40   & 1 & 8.14 & 0.020 & W  \\                              
beq6   &    40   & 4 & 9.85 & 0.071 & W   \\  
beq6   &    40   & 5 & 16.19 & 0.129 & W   \\  

beq7   &    50   & 1 & 16.12 & 0.070 & W  \\                              
beq7   &    50   & 2 & 11.93 & 0.075 & S  \\  

beq8   &    100   & 1 & 12.59 & 0.034 &  S \\                              
beq8   &    100   & 4 & 9.03 & 0.016 &  U  \\  
  
\enddata
\end{deluxetable}

\begin{figure}
\vspace{-1.0in}
\plotone{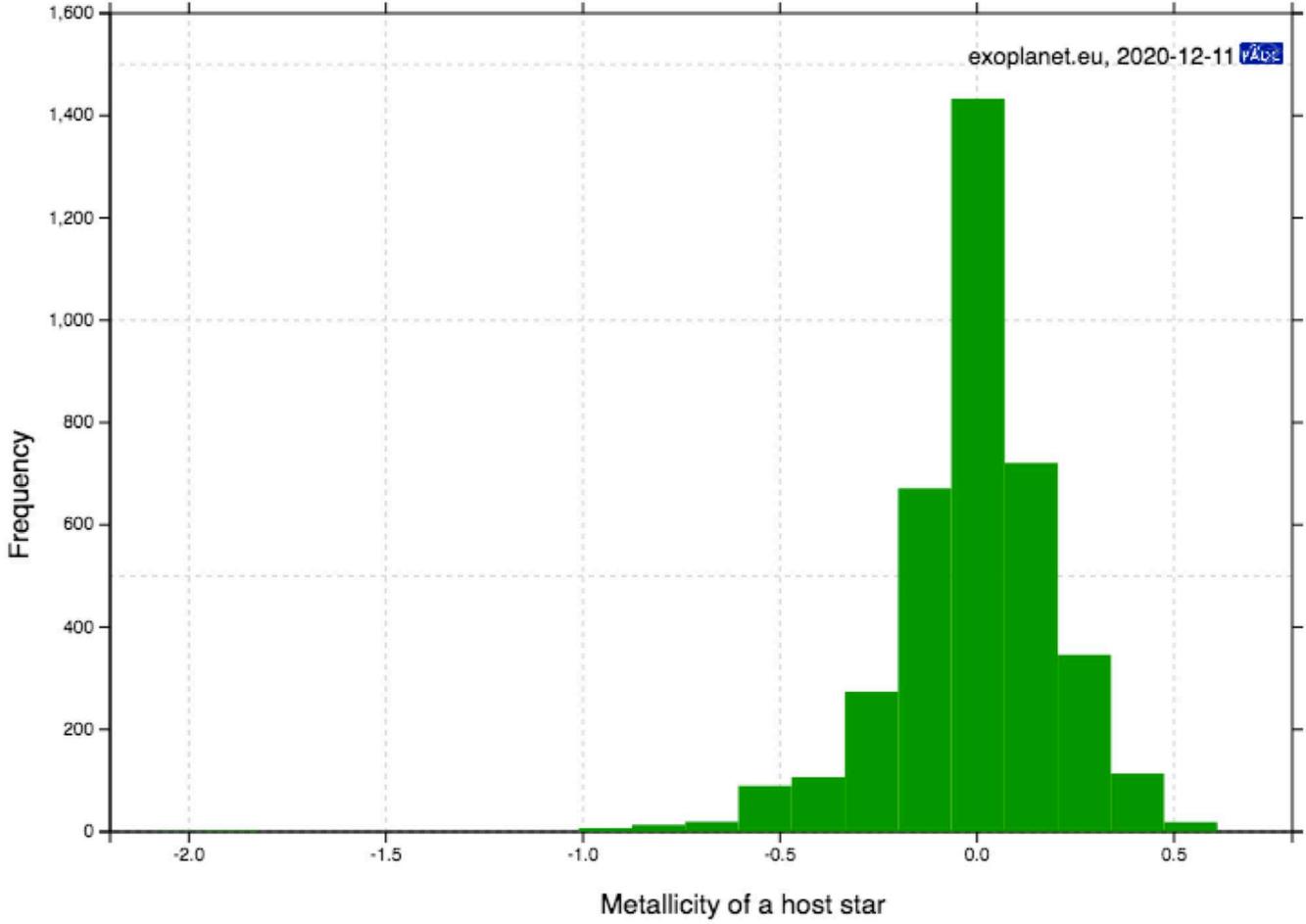}
\vspace{-2.0in}
\caption{Frequency of planets in the Extrasolar Planets Encyclopedia database 
(exoplanet.eu) as a function of host star metallicity. 
Evidently the exoplanet frequency is similar to a Gaussian distribution centered 
on solar metallicity (0.0). }
\end{figure}

\clearpage

\begin{figure}
\vspace{-2.0in}
\plotone{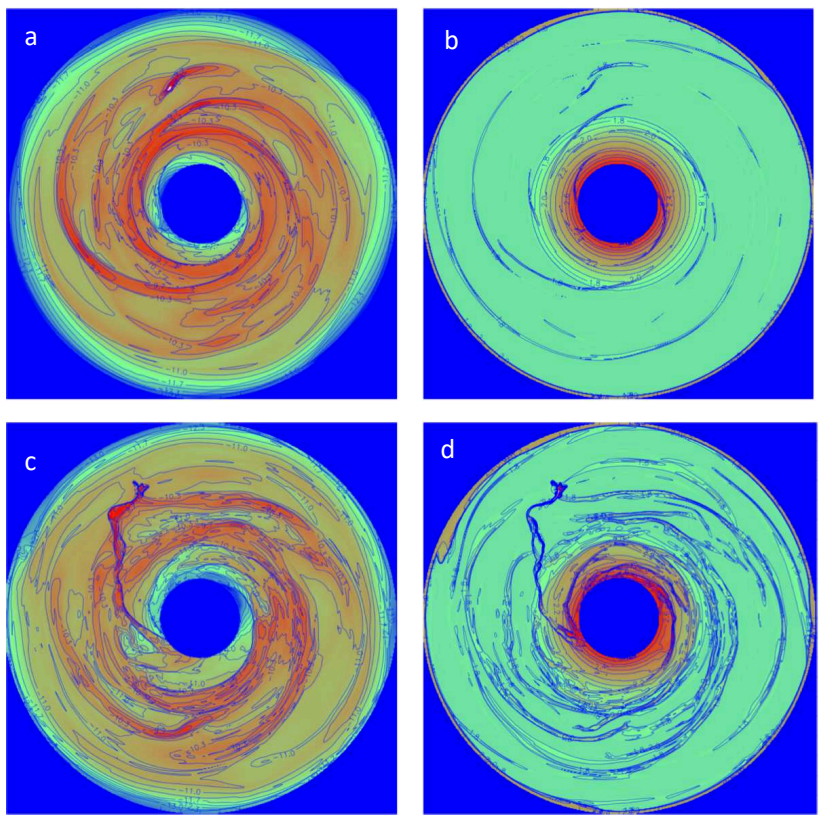}
\vspace{-1.0in}
\caption{Equatorial (midplane) density (left) and temperature (right) contours for 
models beq1 (a,b) at 210 yrs and for model beq3 (c,d) at 223 yrs, respectively.
The disks have inner radii of 4 au and outer radii of 20 au. Density contours are labelled in 
log cgs units and temperature contours are labelled in log K units.
Maximum midplane gas densities are:
(a) $2.5 \times 10^{-9}$ g cm$^{-3}$ and  (c) $7.4 \times 10^{-9}$ g cm$^{-3}$.
The initial maximum midplane density is $1.0 \times 10^{-10}$ g cm$^{-3}$ at 4 au.
The models start with an  initial minimum temperature of 180 K (light orange color) 
and can cool down to a minimum temperature of 40 K (light green color). The red 
color corresponds to temperatures of $\sim$ 1000 K.}
\end{figure}

\clearpage

\begin{figure}
\vspace{-2.0in}
\plotone{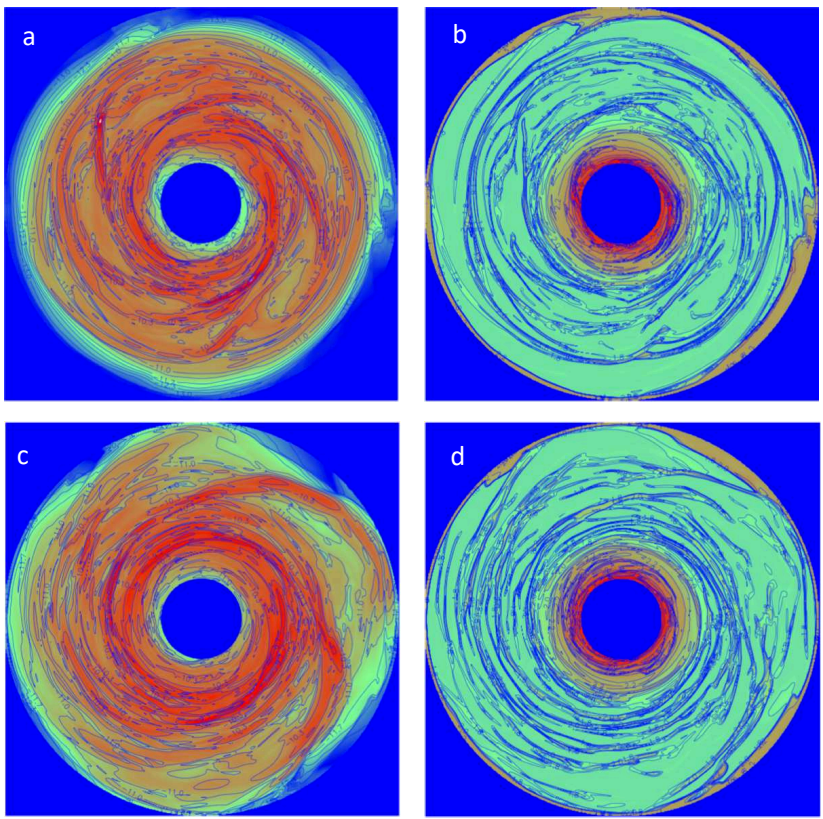}
\vspace{-1.0in}
\caption{Equatorial (midplane) density (left) and temperature (right) contours for 
models beq5 (a,b) at 205 yrs and for model beq8 (c,d) at 326 yrs, respectively,
plotted as in Figure 1. Maximum midplane gas densities are:
(a) $1.2 \times 10^{-9}$ g cm$^{-3}$ and  (c) $6.2 \times 10^{-10}$ g cm$^{-3}$.}
\end{figure}

\clearpage

\begin{figure}
\vspace{-2.0in}
\plotone{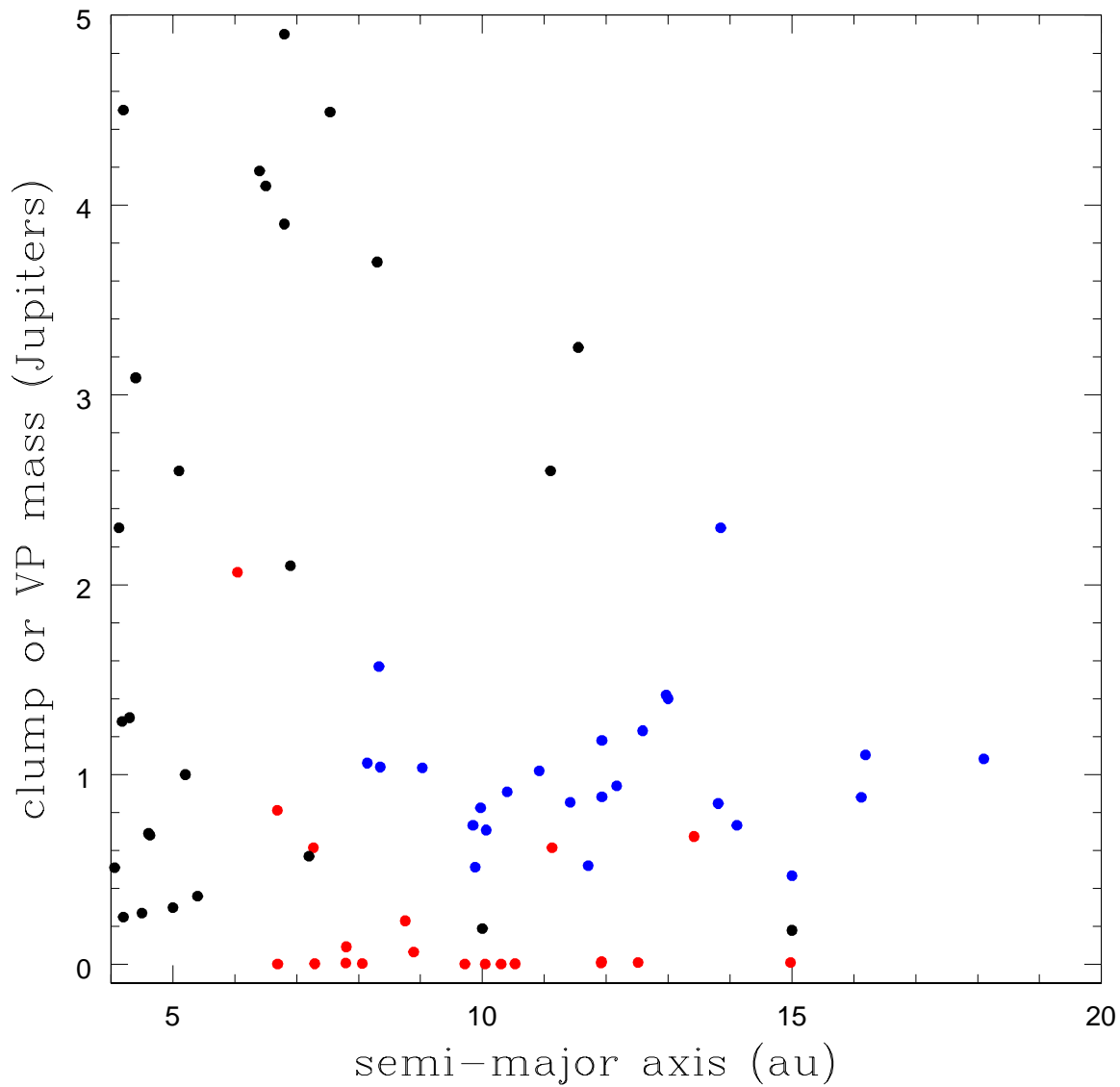}
\caption{Masses and semi-major axes of the S and W clumps from Table 3
are shown in blue, while the red points show the VPs from Boss (2019)
at the data dump closest in time to the final times of
the clumps listed in Table 1. The black points
show all exoplanets listed in the Extrasolar Planets Encyclopedia 
(exoplanets.eu) for masses between 0.1 and 
5 $M_{Jup}$  and semi-major axes between 4 au and 20 au.}
\end{figure}

\clearpage

\begin{figure}
\vspace{-2.0in}
\plotone{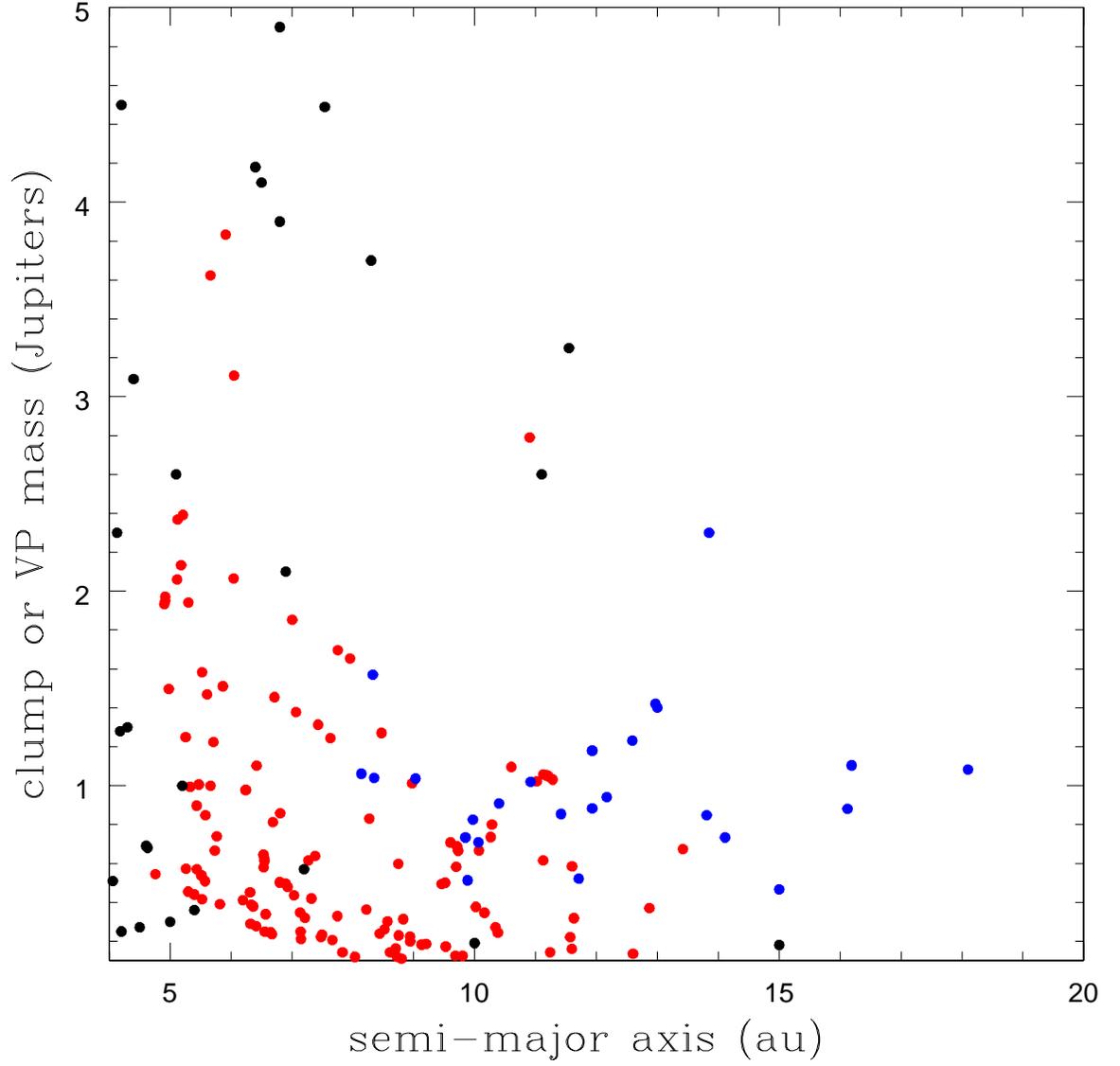}
\caption{Same as Figure 4, except that now the red points show the VPs
from the Boss (2019) models sampled at about 40 times during their evolutions, 
instead of only at the same time as the S and W clumps of the present
models.}
\end{figure}

\clearpage

\begin{figure}
\vspace{-2.0in}
\plotone{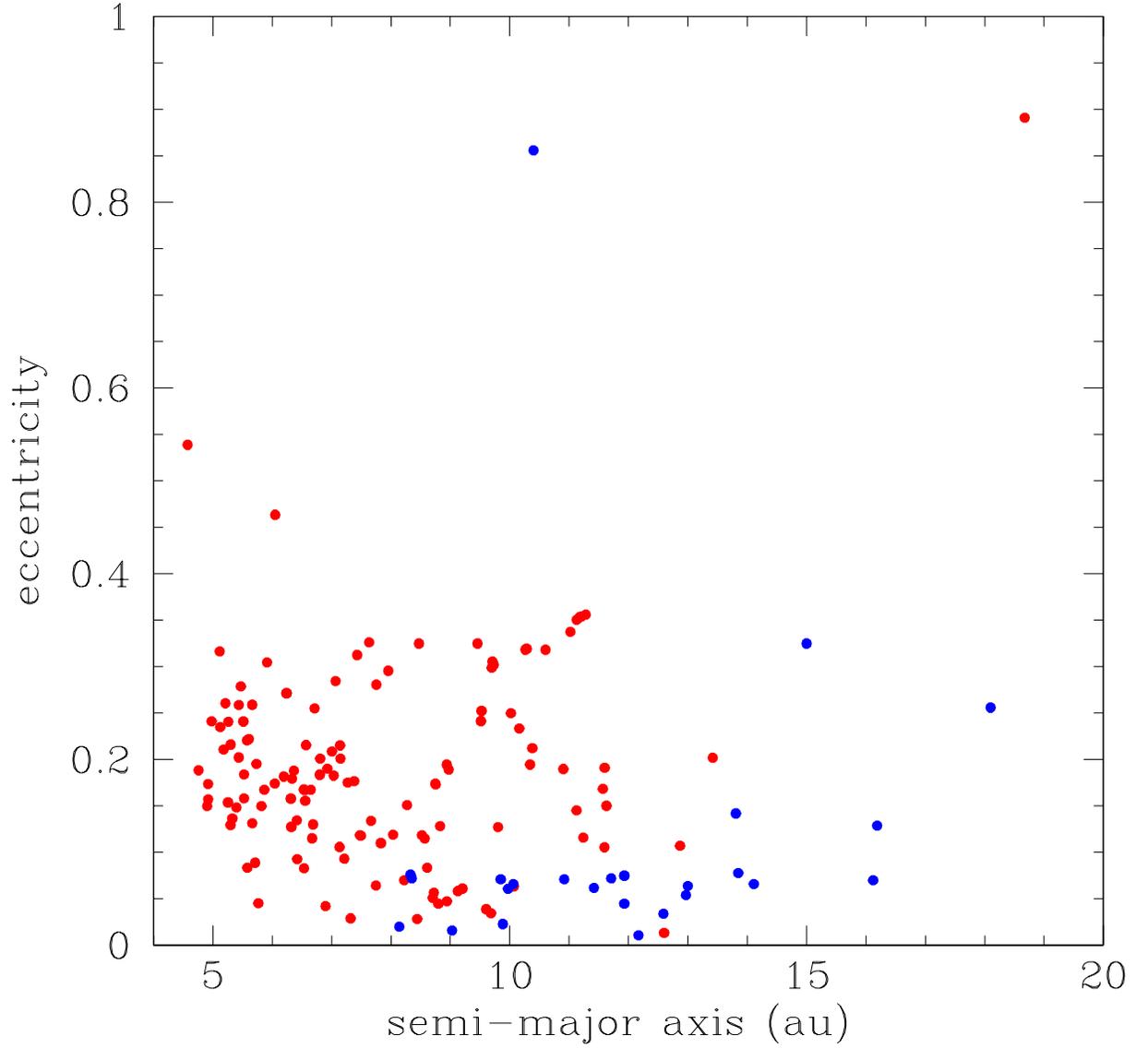}
\caption{Eccentricities and semi-major axes for the clumps listed in Table 3
are shown in blue, while the red points show the values for the VPs
from the Boss (2019) models sampled at about 40 times during their evolutions.}
\end{figure}

\clearpage

\begin{figure}
\vspace{-2.0in}
\plotone{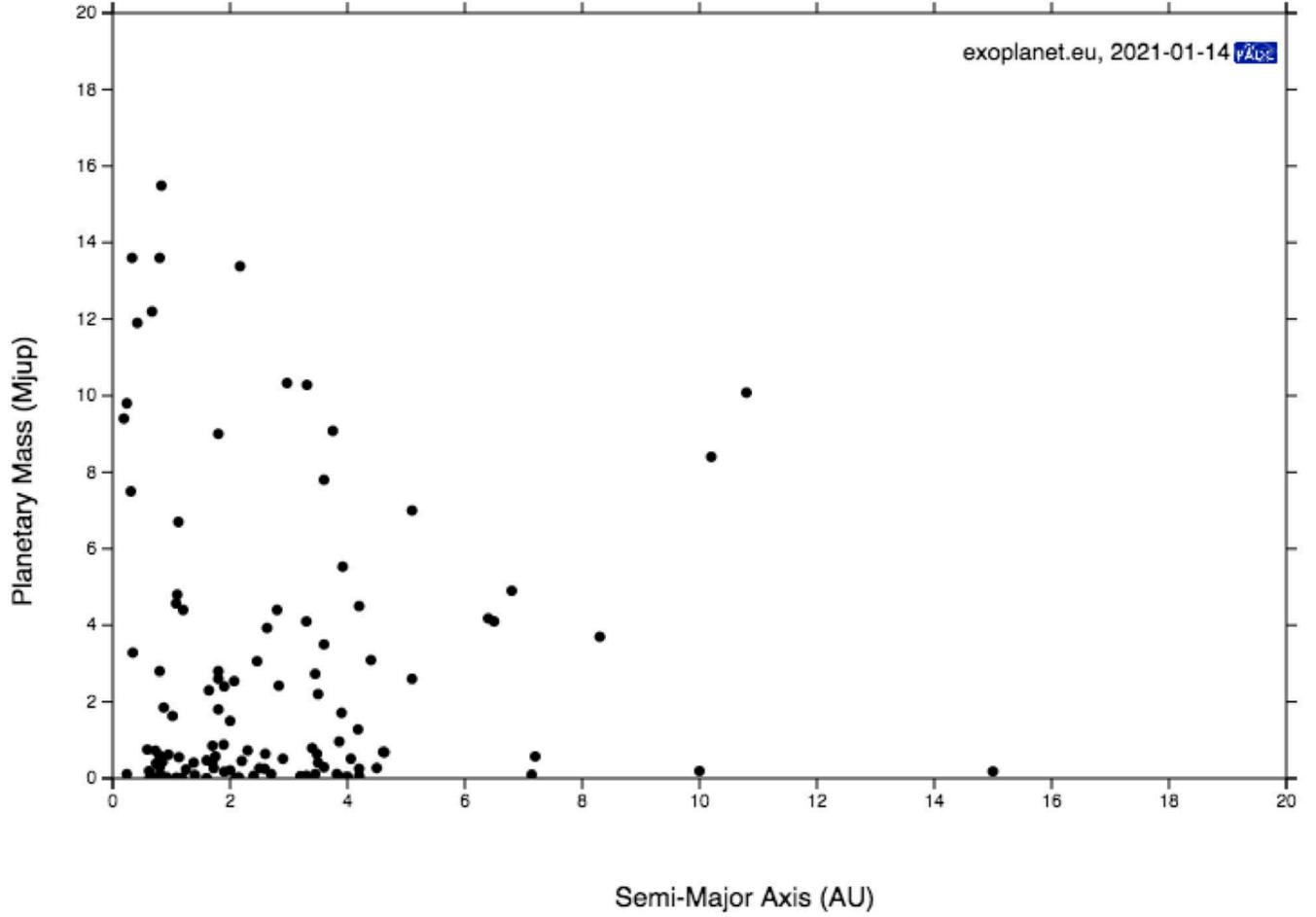}
\caption{Masses and semi-major axes of the confirmed microlensing exoplanets 
listed in the Extrasolar Planets Encyclopedia  (exoplanets.eu) for masses 
between 0.0 and 20 $M_{Jup}$  and semi-major axes between 0 au and 20 au.}
\end{figure}

\end{document}